\documentclass[twocolumn,aps,prd,reprint,superscriptaddress,nofootinbib]{revtex4}

\usepackage{amsmath}
\usepackage{amsfonts}
\usepackage{amssymb}
\usepackage{graphicx}
\usepackage{epsfig}
\usepackage{color}
\usepackage{multirow}
\usepackage[T1]{fontenc} 
\usepackage{todonotes}
\usepackage{hyperref}
\usepackage{listings}

\def\ba{\begin{eqnarray}}
\def\ea{\end{eqnarray}}
\def\bea{\begin{eqnarray}}
\def\eea{\end{eqnarray}}
\def\be{\begin{equation}}
\def\ee{\end{equation}}
\def\d{\mathrm{d}}
\def\({\left(}
\def\){\right)}
\def\[{\left[}
\def\]{\right]}

\def\bi{\begin{itemize}}
\def\ei{\end{itemize}}

\begin{document}


\title{\center{Looking for non-Gaussianity in all the right places: \\ A new basis for non-separable bispectra}}

\author{Joyce Byun}
\email{byun@astro.cornell.edu}
\affiliation{Department of Astronomy, Cornell University, Ithaca, NY 14853, USA}

\author{Nishant Agarwal}
\email{nua11@psu.edu}
\affiliation{Department of Astronomy and Astrophysics, The Pennsylvania State University, University Park, PA 16802, USA}
\affiliation{Institute for Gravitation and the Cosmos, The Pennsylvania State University, University Park, PA 16802, USA}

\author{Rachel Bean}
\email{rbean@astro.cornell.edu}
\affiliation{Department of Astronomy, Cornell University, Ithaca, NY 14853, USA}

\author{Richard Holman}
\email{rh4a@andrew.cmu.edu}
\affiliation{Department of Physics, Carnegie Mellon University, Pittsburgh, PA 15213, USA}

\date{\today}

\begin{abstract}
Non-Gaussianity in the distribution of inflationary perturbations, measurable in statistics of the cosmic microwave background (CMB) and large scale structure fluctuations, can be used to probe non-trivial initial quantum states for these perturbations. The bispectrum shapes predicted for generic non-Bunch-Davies initial states are non-factorizable (``non-separable'') and are highly oscillatory functions of the three constituent wavenumbers. This can make the computation of CMB bispectra, in particular, computationally intractable. To efficiently compare with CMB data one needs to construct a separable template that has a significant similarity with the actual shape in momentum space.  In this paper we consider a variety of inflationary scenarios, with different non-standard initial conditions, and how best to construct viable template matches. In addition to implementing commonly used separable polynomial and Fourier bases, we introduce a basis of localized piecewise spline functions. The spline basis is naturally nearly orthogonal, making it easy to implement and to extend to many modes. We show that, in comparison to existing techniques, the spline basis can provide better fits to the true bispectrum, as measured by the cosine between shapes, for sectors of the theory space of general initial states. As such, it offers a useful approach to investigate non-trivial features generated by fundamental properties of the inflationary Universe.
\end{abstract}

\maketitle


\section{Introduction}
\label{intro}

We are fortunate to live in a time when cosmological datasets can probe the Universe in exquisite detail. In particular, the cosmic microwave background (CMB) provides a rich source of information about the very early Universe, and is an especially precise probe of the inflationary paradigm. An important question that we are now in a position to probe, more thoroughly than ever, is: what is the initial quantum state of inflationary fluctuations? It is usually taken to be the Bunch-Davies state, but from the point of view of treating inflation as an effective theory it is not unreasonable to consider the choice of state to be open, subject to the conditions that it allow for inflation to occur and that it be consistent with field theoretic precepts. Explicit examples of scenarios that give rise to non-Bunch-Davies initial conditions for inflation can be found in \cite{Vilenkin:1982wt,Dey:2011mj,Dey:2012qp,Agullo:2013ai,Lello:2013mfa,Sugimura:2013cra}. Assuming that the initial state is more general than the free vacuum, such as a Bogoliubov transform of the Bunch-Davies state or even a mixed state, we can calculate its imprint on cosmological observables like the power spectrum and bispectrum of inflationary perturbations, and in turn those of CMB temperature anisotropies \cite{Contaldi:1999jr,Martin:2000xs,Kaloper:2002uj,Danielsson:2002kx,Easther:2002xe,Chen:2006nt,Holman:2007na,Meerburg:2009ys,Agullo:2010ws,Ashoorioon:2010xg,Ganc:2011dy,Dey:2011mj,Chialva:2011hc,Kundu:2011sg,Dey:2012qp,Agarwal:2012mq,Brahma:2013rua,Bahrami:2013isa}. Whether these effects can actually be observed in cosmological data depends on the extent of departure from a Bunch-Davies state, the number of e-folds of inflation beyond the minimum required, and of course the sensitivity of our experiments \cite{Flauger:2013hra}.

Given these choices, how can we narrow down the possibilities? The initial state of perturbations will leave its imprints on their correlation functions, for example, logarithmic oscillatory modulations in the CMB power spectrum \cite{Ade:2015lrj}. Higher order correlators, such as the bispectrum \cite{Bennett:2012zja,Ade:2013ydc}, are extremely sensitive to deviations from the Bunch-Davies state. The bispectrum carries information both about the amplitude of the correlations, typically encoded in $f_{\rm NL}$, as well as of preferred configurations in momentum space; the three momenta must form a triangle, but the shape of the triangle is sensitive to both the interactions of the inflaton as well as the initial state. 

The bispectrum for general initial states is highly oscillatory and cannot be written in a separable form, i.e. as a product of separate functions of the three momentum modes. This makes the study of these states via the bispectrum computationally difficult.  For such shapes we usually construct a basis of separable functions and rewrite the desired shape as a sum over many such basis functions, for example, using a polynomial basis \cite{Fergusson:2009nv}, Fourier basis \cite{Meerburg:2010ks}, or divergent basis \cite{Byun:2013jba}.\footnote{In special shape-specific cases, other basis sets can be used: for example, feature or resonant models exhibiting linear or logarithmic oscillations, respectively, can be efficiently reconstructed through a one-dimensional (1D) expansion in the sum of wavenumbers, $k_1+k_2+k_3$ \cite{Munchmeyer:2014nqa,Munchmeyer:2014cca}.}  As long as the original and reconstructed shapes are very similar and have a significant overlap, or in other words have a cosine close to unity, we can look for signatures of the reconstructed shape in the CMB bispectrum and be assured that the result will be an accurate reflection of what we would have obtained for the real shape. Non-Bunch-Davies shapes, however, can be difficult to efficiently describe with existing bases \cite{Fergusson:2009nv,Meerburg:2010ks,Byun:2013jba,Ade:2013ydc}. This leaves open the possibility that signatures of general initial states could still be present, yet undetected, in the CMB data.

In this paper we describe a new basis of piecewise spline functions and use it to fit non-Bunch-Davies shapes of the bispectrum. The {\it spline} basis consists of polynomial functions defined locally, between various points called ``knots'' in 3D space \cite{deBoor:2001,Dierckx:1995}. This makes it particularly suitable for smoothing and interpolating data with complex patterns, and in our case, for capturing localized features of any complicated shape of the bispectrum. Another immediate advantage of defining localized functions is that the basis functions are orthogonal to a good approximation, and there is no need to perform a Gram-Schmidt-like orthogonalization process on the basis functions. This makes the spline basis easy to implement and to extend, using a large number of mode functions to capture fine features in the bispectrum. We find that the spline basis performs at least as well as the polynomial and Fourier bases in describing most non-Bunch-Davies shapes, and for many shapes offers significant improvements (in the cosines) over existing techniques.

The remaining paper is organized as follows. In section \ref{sec:review} we briefly review how we define the initial conditions for the perturbations and obtain the bispectrum for general initial states. We describe the spline basis in section \ref{sec:splines} and use it to fit non-Bunch-Davies shapes in section \ref{sec:splinesNBD}. We conclude in section \ref{sec:discussion} with a summary and discussion on the scope of this work. Appendices \ref{app:powerspectrum} and \ref{app:bispectrum} contain details on calculations of the correlation functions for general initial states, and appendix \ref{app:splinefit} describes our numerical implementation of the spline basis.


\section{Non-Bunch-Davies shapes}
\label{sec:review}

We usually describe primordial correlation functions in terms of the curvature perturbation $\zeta(t,\vec{x})$, since this quantity is conserved outside the horizon \cite{Maldacena:2002vr}. It is defined as the perturbation in the local scale factor $a(t)$; the metric perturbation $h_{ij}(t,\vec{x})$ is then written as $h_{ij} = a^{2} e^{2\zeta} \delta_{ij}$. In an effective field theory setting, $\zeta(t,\vec{x})$ is related to the Goldstone mode of time reparameterization symmetry breaking, usually denoted as $\pi(t,\vec{x})$ \cite{Cheung:2007st,Cheung:2007sv}. In this section we describe how the choice of initial state for the perturbation $\zeta(t,\vec{x})$ (or equivalently $\pi(t,\vec{x})$) affects the primordial bispectrum $B_{\zeta}(k_1, k_2, k_3)$ defined via $\big\langle \zeta_{\vec{k}_1} \zeta_{\vec{k}_2} \zeta_{\vec{k}_3} \big\rangle = (2\pi)^{3} \delta^3 \big( \vec{k}_1 + \vec{k}_2 + \vec{k}_3 \big) B_{\zeta} (k_1, k_2, k_3)$.

Starting with the Einstein-Hilbert action, we can calculate the action for scalar perturbations directly in the $\zeta$-gauge. Writing down the most general Lorentz-invariant scalar-tensor theory with second order equations of motion results in the Horndeski action for the perturbations \cite{Horndeski:1974wa,Gao:2011qe,DeFelice:2011uc,Burrage:2011hd,Ribeiro:2012ar}.\footnote{The Horndeski action does not describe ghost inflation, however, which can be included in an effective field theory setting. Another example outside the Horndeski domain is Hor\v{a}va-Lifshitz gravity, in which Lorentz invariance is explicitly broken.} For $P(X,\phi)$ models of inflation, with $X \equiv -g_{\mu\nu} \partial^{\mu}\phi \partial^{\nu}\phi$,\footnote{The action for many single scalar field models of inflation can be written as $S = \frac{1}{2} \int \d^4 x \sqrt{-g} \[ R + 2P(X,\phi) \]$, with $\phi$ controlling the dynamics of both the background and perturbations.} the leading order in slow-roll Horndeski action at cubic order in the perturbations is given by
\bea
	S & = & \int \d^3 x \, \d t \, a^3 \Bigg\{ \frac{\varepsilon}{c_s^2} \[ \dot{\zeta}^2 - \frac{c_s^2}{a^2} (\partial_i \zeta)^2 \] + \Lambda_1 \dot{\zeta}^3 + \Lambda_2 \zeta \dot{\zeta}^2 \nonumber \\
	& & \quad \quad \quad + \ \frac{\Lambda_3}{a^2} \zeta(\partial_i\zeta)^2 \Bigg\} \, + \, S_{\rm boundary} \, .
\label{eq:action}
\eea
Here we have set $M_{\rm Pl} = 1$, $\varepsilon \equiv -\dot{H}/H^2$ is the slow-roll parameter, $c_s$ is the effective sound speed for perturbations, and the couplings $\Lambda_1 - \Lambda_3$ are given by
\bea
	\Lambda_1 & = & \frac{\varepsilon}{Hc_s^4} \( 1 - c_s^2 - 2\frac{\lambda c_s^2}{\Sigma} \) \, , \\
	\Lambda_2 & = & -3 \frac{\varepsilon}{c_s^4} \( 1 - c_s^2 \) \, , \\
	\Lambda_3 & = & \frac{\varepsilon}{c_s^2} \( 1 - c_s^2 \) \, ,
\eea
with $\lambda \equiv X^2 \frac{\partial^2 P}{\partial X^2} + \frac{2}{3} X^3 \frac{\partial^3 P}{\partial X^3}$ and $\Sigma \equiv X \frac{\partial P}{\partial X} + 2 X^2 \frac{\partial^2 P}{\partial X^2} = \frac{H^2\varepsilon}{c_s^2}$. The boundary terms in eq.\ (\ref{eq:action}) are important and in general do contribute to the bispectrum when the initial state is different from the Bunch-Davies vacuum; here we assume for simplicity that the initial state for $\zeta(t,\vec{x})$ is defined in such a way as to cancel these boundary terms. It is also worth noting that the action in eq.\ (\ref{eq:action}) is equivalent to that obtained using an effective field theory approach \cite{Cheung:2007st,Cheung:2007sv} up to boundary terms, and the bispectrum for general initial states agrees between the two actions, as shown in \cite{Agarwal:2013rva}.

The usual method to obtain correlation functions of $\zeta(t,\vec{x})$ is to use in-in perturbation theory \cite{Schwinger:1960qe,Mahanthappa:1962ex,Bakshi:1962dv,Kadanoff:1962,Bakshi:1963bn,Keldysh:1964ud,Jordan:1986ug,Calzetta:1986ey}. The initial state can be input as a density matrix defined at the initial time $t_0$ \cite{Agarwal:2012mq}; for simplicity we will assume that the initial density matrix is pure and Gaussian.\footnote{Here by ``pure'' we are distinguishing between pure and mixed quantum states. Mathematically speaking, ${\rm Tr} \( \rho^2 \) = 1$ for pure states, while ${\rm Tr} \( \rho^2 \) < 1$ for mixed states, $\rho$ being the density matrix. Relaxing the pure state assumption leads to qualitatively very similar results to what we discuss here. By ``Gaussian'' we mean that the action describing the initial state is quadratic.} Doubling the fields on the plus and minus branches of the in-in contour, we can calculate the Green's function corresponding to the quadratic (or ``free'') part of the action in eq.\ (\ref{eq:action}), including the effect of a general initial state; details of this calculation can be found in the appendix of \cite{Agarwal:2012mq}. The Green's function for $\zeta(t,\vec{x})$ is found to be
\bea
	{\cal G}_k^{\zeta}(\eta,\eta') & = & \frac{c_s^2}{2\varepsilon} \frac{1}{a(\eta)a(\eta')}
	\left(
	\begin{array}{c c}
		G^{++}_{k} (\eta, \eta') & G^{+-}_{k} (\eta, \eta') \\
		G^{-+}_{k} (\eta, \eta') & G^{--}_{k} (\eta, \eta')
	\end{array}
	\right) \, , \nonumber \\
\eea
where $\eta$ is the conformal time defined as $\eta = \int \d t/a$ and the factor out front comes from rewriting the quadratic action in terms of the canonically rescaled field $\chi(t,\vec{x})$, $\zeta = \frac{1}{\sqrt{2\varepsilon}} \frac{c_s}{a} \chi$. The functions $G_k^{\pm,\pm} (\eta, \eta')$ are given by
\bea
	G_{k}^{++}(\eta,\eta') & = & f_k^{>}(\eta) f_k^{<}(\eta') \theta (\eta - \eta') \nonumber \\
	& & \quad + \ f_k^{<}(\eta) f_k^{>}(\eta') \theta (\eta' - \eta) \, ,
\label{eq:Gkpp} \\
	G_{k}^{+-}(\eta,\eta') & = & f_k^{<}(\eta) f_k^{>}(\eta') \, , \\
	G_{k}^{-+}(\eta,\eta') & = & G_{k}^{+-*}(\eta,\eta') \, , \\
	G_{k}^{--}(\eta,\eta') & = & G_{k}^{++*}(\eta,\eta') \, ,
\eea
and the mode functions $f_k^{\gtrless}(\eta)$ are solutions to the second order differential equation resulting from the Green's function equation,
\bea
	f_k^{\gtrless}(\eta) & = & \alpha_k^{\gtrless} h_k(\eta) + \beta_k^{\gtrless} h^{*}_k(\eta) \nonumber \\
	& & \quad - \ 2f_k^{\gtrless}(\eta_0) A_k g_k(\eta) \theta(\eta_0 - \eta) \, .
\label{eq:fkgtrless}
\eea
The last term above is an additional contribution from the initial density matrix, $A_k$ being the kernel that multiplies the $\zeta^{+}_{\vec{k}} \zeta^{+}_{-\vec{k}}$ term in the initial state action \cite{Agarwal:2012mq}; it does not, however, contribute to the bispectrum, and so we will ignore it in our discussion. The Bogoliubov coefficients $\alpha_k^{\gtrless}$, $\beta_k^{\gtrless}$ are functions of kernels in the initial density matrix, the mode functions, and time derivatives of the mode functions, all at the initial time, and satisfy $\alpha_{k}^{<} = \beta_{k}^{>*}$, $\beta_{k}^{<} = \alpha_{k}^{>*}$, $\left |\alpha_k^{>} \right|^2 - \left |\beta_k^{>} \right|^2 = 1$. Finally, the function $h_k(\eta)$ is defined at leading order in slow-roll as
\bea
	h_k(\eta) & = & -\frac{1}{2} (-\pi\eta)^{1/2} H_{3/2}^{(1)} (-c_s k \eta) \, ,
\label{eq:hk}
\eea
where $H_{3/2}^{(1)}$ is a Hankel function. The Bunch-Davies choice consists of setting the initial time $\eta_0 \rightarrow -\infty$, the initial density matrix to unity, and additionally $\alpha_k^{>} = 1$, $\beta_k^{>} = 0$ so that the mode function $f_k^{>}(\eta)$ picks out the positive frequency solution proportional to $e^{-i c_s k \eta}$.

The time $\eta_0$ at which the initial conditions are set can be taken to be a constant time in the past at the onset of infation, or can be considered as a scale-dependent quantity, $\eta_0(k)$. In the latter case, the initial conditions for each $k$ mode are set at the time when the physical momentum corresponding to this mode $c_s k/a(\eta_0)$ (with $a(\eta_0) = -1/(\eta_0 H$) during inflation, at leading order) crosses a fixed energy scale $\Lambda$ of new physics. The Bogoliubov transform in eq.\ (\ref{eq:fkgtrless}) is correct for either choice of $\eta_0$. In appendix \ref{app:powerspectrum} we show that for a scale-dependent initial time, this solution leads to the well-known oscillations in the late-time power spectrum \cite{Danielsson:2002kx,Easther:2002xe}. We use both choices of initial time in section \ref{sec:splinesNBD} when we apply the spline basis to non-Bunch-Davies shapes of the bispectrum.

Let us now discuss how general initial states modify the bispectrum. Observables, such as the bispectrum, can be calculated using any combination of plus and minus fields on the in-in contour. For Gaussian initial states, we can calculate the three-point function in the perturbations as
\bea
	\left\langle \zeta_{\vec{k}_1}^{+} \zeta_{\vec{k}_2}^{+} \zeta_{\vec{k}_3}^{+} \right\rangle (\eta) & = & \Big\langle \zeta_{\vec{k}_1}^{+}(\eta) \zeta_{\vec{k}_2}^{+}(\eta) \zeta_{\vec{k}_3}^{+}(\eta) \nonumber \\
	& & \ \ \times \, \exp \left[i\left(S^{(3)+} - S^{(3)-}\right)\right] \Big\rangle_{\rm G} \, , \quad \quad
\label{eq:threepointdef}
\eea
where $S^{(3)}$ is the cubic part of the action in eq.\ (\ref{eq:action}) written in momentum space, with conformal time derivatives, and with the time integral running from $\eta_0$ to $\eta$. The subscript ``G'' indicates that Wick contractions on the right are carried out using the Gaussian theory. At leading order, only the three operators $\dot{\zeta}^3$, $\zeta\dot{\zeta}^2$, and $\zeta(\partial_i\zeta)^2$ contribute, and we can write the three-point function at late times $\left\langle \zeta_{\vec{k}_{1}}^{+} \zeta_{\vec{k}_{2}}^{+} \zeta_{\vec{k}_{3}}^{+} \right\rangle (\eta) \big|_{\eta \rightarrow 0^{-}}$ as a sum of contributions from each of these three operators. To calculate the three-point function in eq.\ (\ref{eq:threepointdef}) at late times we need the function $G_k^{\zeta,++}(0,\eta')$ and its derivative $\partial_{\eta'} G_k^{\zeta,++}(0,\eta')$ for $\eta' \geq \eta_0$; using eqs.\ (\ref{eq:Gkpp}), (\ref{eq:fkgtrless}) (discarding the $\theta(\eta_0 - \eta)$ term), and (\ref{eq:hk}) these are given by
\bea
	G_k^{\zeta,++}(0,\eta') & = & \frac{H^2}{4\varepsilon c_{s}k^{3}} \Big[ a_{k} (1-ic_{s}k\eta') e^{ic_{s}k\eta'} \nonumber \\
	& & \quad \quad + \ b_{k} (1+ic_{s}k\eta') e^{-ic_{s}k\eta'} \Big]
\eea
and
\bea
	\partial_{\eta'} G_{k}^{\zeta,++}(0,\eta') & = & \frac{H^2 c_{s}\eta'}{4\varepsilon k}  \left( a_{k} e^{ic_{s}k\eta'} + b_{k} e^{-ic_{s}k\eta'} \right) \, , \nonumber \\
\eea
where we have defined the functions
\bea
	a_{k} & = & \( \alpha_{k}^{>} - \beta_{k}^{>} \) \alpha_{k}^{>*} \, , \\
	b_{k} & = & - \( \alpha_{k}^{>} - \beta_{k}^{>} \) \beta_{k}^{>*} \, .
\eea
Using these in eq.\ (\ref{eq:threepointdef}) and performing the time integrals we find that the contributions to the bispectrum from the three operators are given by
\bea
	& & \left\langle \zeta_{\vec{k}_{1}}^{+} \zeta_{\vec{k}_{2}}^{+} \zeta_{\vec{k}_{3}}^{+} \right\rangle_{\dot{\zeta}^3} (\eta) \big|_{\eta \rightarrow 0^{-}} \ = \ -\frac{3}{32} \(2\pi\)^{3} \delta^{3}\(\sum \vec{k}_{i}\) \nonumber \\
	& & \quad \times \ \Lambda_1 \frac{H^{5}}{\varepsilon^{3}} \Bigg[ \sum_{l,m,n = 0}^{1} c_{k_{1}}^{(l)} \ c_{k_{2}}^{(m)} \ c_{k_{3}}^{(n)} \nonumber \\
	& & \quad \quad \times \ {\cal F}_{\dot{\zeta}^3} \left( (-1)^{l}k_{1}, (-1)^{m}k_{2}, (-1)^{n}k_{3},\eta_{0} \right) \Bigg] \nonumber \\
	& & \quad + \ {\rm c.c.} \, ,
\label{eq:bispec1}
\eea
\bea
	& & \left\langle \zeta_{\vec{k}_{1}}^{+} \zeta_{\vec{k}_{2}}^{+} \zeta_{\vec{k}_{3}}^{+} \right\rangle_{\zeta\dot{\zeta}^2} (\eta) \big|_{\eta \rightarrow 0^{-}} \ = \ \frac{1}{32} \(2\pi\)^{3} \delta^{3}\(\sum \vec{k}_{i}\) \nonumber \\
	& & \quad \times \ \Lambda_2 \frac{H^{4}}{\varepsilon^{3}} \Bigg[ \sum_{l,m,n = 0}^{1} c_{k_{1}}^{(l)} \ c_{k_{2}}^{(m)} \ c_{k_{3}}^{(n)} \nonumber \\
	& & \quad \quad \times \ {\cal F}_{\zeta\dot{\zeta}^2} \left( (-1)^{l}k_{1}, (-1)^{m}k_{2}, (-1)^{n}k_{3},\eta_{0} \right) \Bigg] \nonumber \\
	& & \quad + \ {\rm c.c.} \, ,
\label{eq:bispec2}
\eea
and
\bea
	& & \left\langle \zeta_{\vec{k}_{1}}^{+} \zeta_{\vec{k}_{2}}^{+} \zeta_{\vec{k}_{3}}^{+} \right\rangle_{\zeta(\partial\zeta)^2} (\eta) \big|_{\eta \rightarrow 0^{-}} \ = \ \frac{1}{64} \(2\pi\)^{3} \delta^{3}\(\sum \vec{k}_{i}\) \nonumber \\
	& & \quad \times \ \Lambda_3 \frac{H^{4}}{c_s^2\varepsilon^{3}} \Bigg[ \sum_{l,m,n = 0}^{1} c_{k_{1}}^{(l)} \ c_{k_{2}}^{(m)} \ c_{k_{3}}^{(n)} \nonumber \\
	& & \quad \quad \times \ {\cal F}_{\zeta(\partial\zeta)^2} \left( (-1)^{l}k_{1}, (-1)^{m}k_{2}, (-1)^{n}k_{3},\eta_{0} \right) \Bigg] \nonumber \\
	& & \quad + \ {\rm c.c.} \, ,
\label{eq:bispec3}
\eea
where ``c.c.'' denotes complex conjugate and
\bea
	c_{k}^{(i)} & = &
	\left\{
		\begin{array}{c c}
			a_{k} \ \ & i = 0 \\
			b_{k} \ \ & \ \, i = 1 \, .
		\end{array}
	\right.
\eea
The functions ${\cal F}_{\dot{\zeta}^3}$, ${\cal F}_{\zeta\dot{\zeta}^2}$, and ${\cal F}_{\zeta(\partial\zeta)^2}$ are written out explicitly in appendix \ref{app:bispectrum}. The above equations give us the leading order result for the bispectrum for general initial states. The non-Bunch-Davies contributions to the bispectrum are strongly peaked in the flattened limit $k_1 \approx k_2 + k_3$ (assuming that $k_1$ is the largest momentum mode) and in the squeezed limit $k_3 \ll k_1 \approx k_2$. (We show these enhancements and discuss apparent divergences in both limits in appendix \ref{app:bispectrum}.) Further, these shapes are highly oscillatory, which makes them even harder to constrain using the CMB bispectrum. In the next section we discuss the spline basis that we use to rewrite these shapes as a sum of separable functions.


\section{The spline basis}
\label{sec:splines}

\begin{figure}[!t]
\centering
	\includegraphics[width=0.45\textwidth]{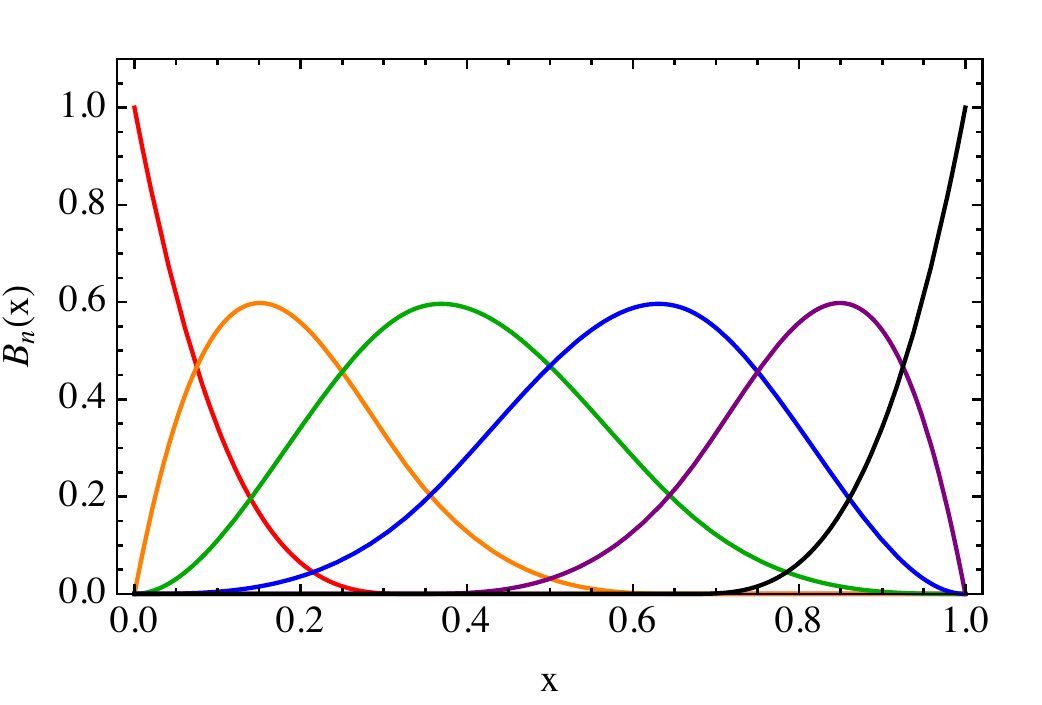}
	\caption{Example spline basis generated from knots at $x = \{ 0,0,0,0,1/3,2/3,1,1,1,1 \}$ and polynomials of degree $q = 3$.}
\label{fig:sample_1d_bsplines}
\end{figure}

B-splines, short for ``basis splines'', are a well-established, and conceptually simple, mathematical formalism for curve fitting, using a set of piecewise polynomial functions \cite{deBoor:2001,Dierckx:1995}. For a basis in one dimension, one chooses a set of ``knots'', $\{x_0, x_1, \dots, x_N\}$, representing the points at which the polynomial function pieces are joined, and the degree $q$ of the polynomials. For example, a 1D spline basis spanning $0 \leq x \leq 1$ with a set of six piecewise cubic polynomials is shown in fig.\ \ref{fig:sample_1d_bsplines}.  As an explicit example of the functional form of the basis, one of the basis functions shown in the figure is
\bea
	B_1(x) & = &
	\left\{
		\begin{array}{c c}
			\frac{9}{4} \( 4x-18x^2+21x^3 \) \ \ & 0 \leq x < \frac{1}{3} \vspace{2.5pt} \\
			\frac{1}{4} \( 8-36x+54x^2-27x^3 \) \ \ & \ \frac{1}{3} \leq x \leq \frac{2}{3} \, .
		\end{array}
	\right. \nonumber \\
\eea
The numerical coefficients defining the spline basis for input knots and $q$ are easily generated using existing software libraries in many languages.

A general 1D function, $f(x)$, can  be expanded, and approximated, using the spline basis,
\bea
	f'(x) & = & \sum_{n = 0}^{N-1} \alpha_n B_n(x) \, .
\eea
The expansion coefficients are computed by first tabulating a sample of $(x_i, f_i)$ values, where $f_i \equiv f(x_i)$, corresponding to $M$ data points, and finding the values of $\{\alpha_n\}$ that minimize the least-squares function,
\bea
	{\rm LS} & = & \sum_{i=1}^M \left( f_i - \sum_{n=0}^{N-1} \alpha_n B_n(x_i) \right)^2 \, .
\eea
In practice, this requires solving the linear system of equations given by $B^T B \vec{\alpha} = B^T \vec{f}$, where $B$ here is an $M \times N$ matrix containing the values of $B_n(x_i)$.

Similarly, we can generate a higher-dimensional basis by multiplying 1D b-splines together. For example, a 3D b-spline fit for a shape\footnote{A shape function is usually denoted by $S(k_1,k_2,k_3)$, not to be confused with the action $S$.} can be expressed as
\bea
	S'(k_1,k_2,k_3) & = & \sum_{k=0}^{N-1} \sum_{j=0}^k \sum_{i=0}^j A_{ijk} \big[ B_i(k_1) B_j(k_2) B_k(k_3) \nonumber \\
	& & \quad \quad + \ \mathrm{perms} \big] \, ,
\eea
where ``perms'' refers to permutations of $k_1$, $k_2$, and $k_3$. We note that the number of relevant $(i,j,k)$ combinations in such a fit is not $N^3$, because each spline mode is symmetric in the three wavenumbers, and not all $(i,j,k)$ combinations will correspond to a spline mode that is non-zero for wavenumber combinations that can form a closed triangle. In the 2D and 3D cases, efficient algorithms solving for the expansion coefficients $A_{ijk}$ already exist in the literature \cite{Eilers:2006}. We illustrate how to use the spline basis further through explicit examples and include snippets of our numerical codes in appendix \ref{app:splinefit}.
 
In contrast to other bases used to create non-Gaussian templates made of globally varying functions, such as polynomial, Fourier, and divergent functions, the spline basis consists of a set of localized modes, each of which describes only a small region of the allowed $k$-space. This makes the spline basis very well-suited to describing shapes such as those of non-Bunch-Davies models, that are characterized by highly-peaked features concentrated on very flattened triangles. Because of the localization of spline modes, the modes are by construction nearly orthogonal, and no orthogonalization procedure (such as Gram-Schmidt) is used in our implementation.  Avoiding any explicit orthogonalization is an advantage, as orthogonalizing the polynomial/Fourier basis sets via Gram-Schmidt is numerically unstable, and requires very high precision throughout to create a basis with a large number of modes.\footnote{The numerical instability of the classical Gram-Schmidt algorithm can be partially mitigated by instead adopting the modified Gram-Schmidt algorithm, which is what we have implemented in generating the polynomial and Fourier modes.}


\section{Fitting non-Bunch-Davies shapes}
\label{sec:splinesNBD}

In this section we consider two non-Bunch-Davies shapes included in Planck's analysis that were not well-reconstructed in multipole-space using the polynomial modal expansion, $S_{{\rm NBD}1}$ and $S_{{\rm NBD}2}$ \cite{Ade:2013ydc}.\footnote{In the notation of \cite{Ade:2013ydc}, $S_{{\rm NBD}i}$ here is equal to $(k_1 k_2 k_3)^2 B_{\Phi}^{{\rm NBD}i}/\( 2A^2 f_{\rm NL}^{{\rm NBD}i} \)$, where $i = 1 \, , 2$.} We also consider a set of generalized non-Bunch-Davies shapes, $S_{\dot{\zeta}^3}$, $S_{\zeta\dot{\zeta}^2}$, and $S_{\zeta(\partial\zeta)^2}$,\footnote{The shapes $S_{\dot{\zeta}^3}$, $S_{\zeta\dot{\zeta}^2}$, and $S_{\zeta(\partial\zeta)^2}$ here are related to corresponding leading order non-Bunch-Davies corrections to the bispectrum; for example, $S_{\dot{\zeta}^3}(k_1, k_2, k_3) = (k_1 k_2 k_3)^2 \Big[ {\cal F}_{\dot{\zeta}^3}(-k_1, k_2, k_3, \eta_0) + {\cal F}_{\dot{\zeta}^3}(k_1, -k_2, k_3, \eta_0) + {\cal F}_{\dot{\zeta}^3}(k_1, k_2, -k_3, \eta_0) + {\rm c.c.} \Big]$, where we have set $a_{k_i} = 1$ and $b_{k_i} = 0.01$. In general $b_{k_i}$ can have some scale-dependence as long as it does not spoil constraints from backreaction of the energy density in the initial state.} with different assumptions for the initial time boundary $c_s\eta_0$ and its potential wavenumber dependence, $c_s |\eta_0| = 10^3$ Mpc, $(\Lambda/H)/(k_1+k_2+k_3)$, and $(\Lambda/H)/k_1$, where we allow $\Lambda/H$ to be 10 or $10^3$, representing a physically-motivated range of values. These shapes span a wide variety of possible non-Bunch-Davies features in the bispectrum, and allow us to compare our results with previous analyses.

For any non-separable primordial shape $S$, and given a choice of basis $\{M_n\}$, one can compute a separable fit $S'$ that approximates $S$ as a linear combination of separable basis functions,
\bea
	S'(k_1,k_2,k_3) & = & \sum_n \alpha_n \mathcal{M}_n(k_1,k_2,k_3) \, .
\eea
The similarity between the original shape and its fit is quantified by a cosine,
\bea
	\cos(S,S') & \equiv & \frac{\left<S,S'\right>}{\sqrt{\left<S,S\right>\left<S',S'\right>}} \, ,
\eea
where the inner product is defined in Fourier space with a choice of weighting,
\bea
	\left<S,S'\right> & \equiv & \int \d\mathcal{V}_T \, S(k_1,k_2,k_3) S'(k_1,k_2,k_3) w(k_1,k_2,k_3) \, . \nonumber \\
\eea
The volume $\mathcal{V}_T$ includes only those combinations of $k_1$, $k_2$, and $k_3$ that can form a closed triangle, with each wavenumber satisfying $k_{\rm min} \leq k_1, \, k_2, \, k_3 \leq k_{\rm max}$, where $k_{\rm min} = 10^{-3} \; {\rm Mpc}^{-1}$ and $k_{\rm max} = 0.1 \; {\rm Mpc}^{-1}$. The weight $w(k_1,k_2,k_3)$ is typically taken to be either unity or $1/(k_1+k_2+k_3)$, where the latter choice is meant to represent a more accurate reflection of the scaling of the covariance of the CMB bispectrum, such that the Fourier-space cosine is closer to the multipole-space cosine between the CMB bispectra corresponding to the shapes $S$ and $S'$. However, in general we find that both choices result in similar cosines. In our analysis, we use a unit weight for the 3D Fourier basis fits and all 2D fits, and a weight of $1/(k_1+k_2+k_3)$ for the remaining fits.

In this section we implement the existing polynomial and Fourier basis methods, and additionally our new basis of piecewise splines, to obtain separable fits to non-Bunch-Davies shapes. In each case, we quantify the performance of the basis by computing the cosine as a function of an increasing number of modes used in the fit.

We use the polynomial basis described in \cite{Fergusson:2009nv} and the Fourier basis described in \cite{Meerburg:2010ks}. In each case, three 1D functions based on either polynomials or sines/cosines of $k_i$ are multiplied together to form 3D separable functions, that are then orthogonalized using a Gram-Schmidt algorithm to create a basis of 3D orthonormalized and separable functions, called $\{ \mathcal{R}_n \}$ for polynomials and $\{ \mathcal{F}_n \}$ for Fourier modes. Since the basis functions are orthonormal in each case, the expansion coefficients $\{\alpha_n\}$ can be computed through inner products between $S$ and the basis functions, $\alpha_n = \left<S,\mathcal{R}_n\right>$ or $\alpha_n = \left<S,\mathcal{F}_n\right>$.

Alternatively, in the spline basis expansion, three 1D piecewise spline functions are multiplied together to form a basis of 3D separable spline functions $\{B_n\}$. In this case, each mode is highly localized in a region of Fourier space, so any two modes are orthogonal by construction unless they have peaks that overlap. While the lack of strict orthogonality means that the expansion coefficients in the spline basis cannot be computed using simple inner products, existing algorithms can solve for the coefficients efficiently \cite{Eilers:2006}. 

In the subsections that follow, we present polynomial, Fourier, and spline basis fits to a variety of non-Bunch-Davies shapes.

\begin{figure*}[!t]
\centering
	\includegraphics[width=1.0\textwidth]{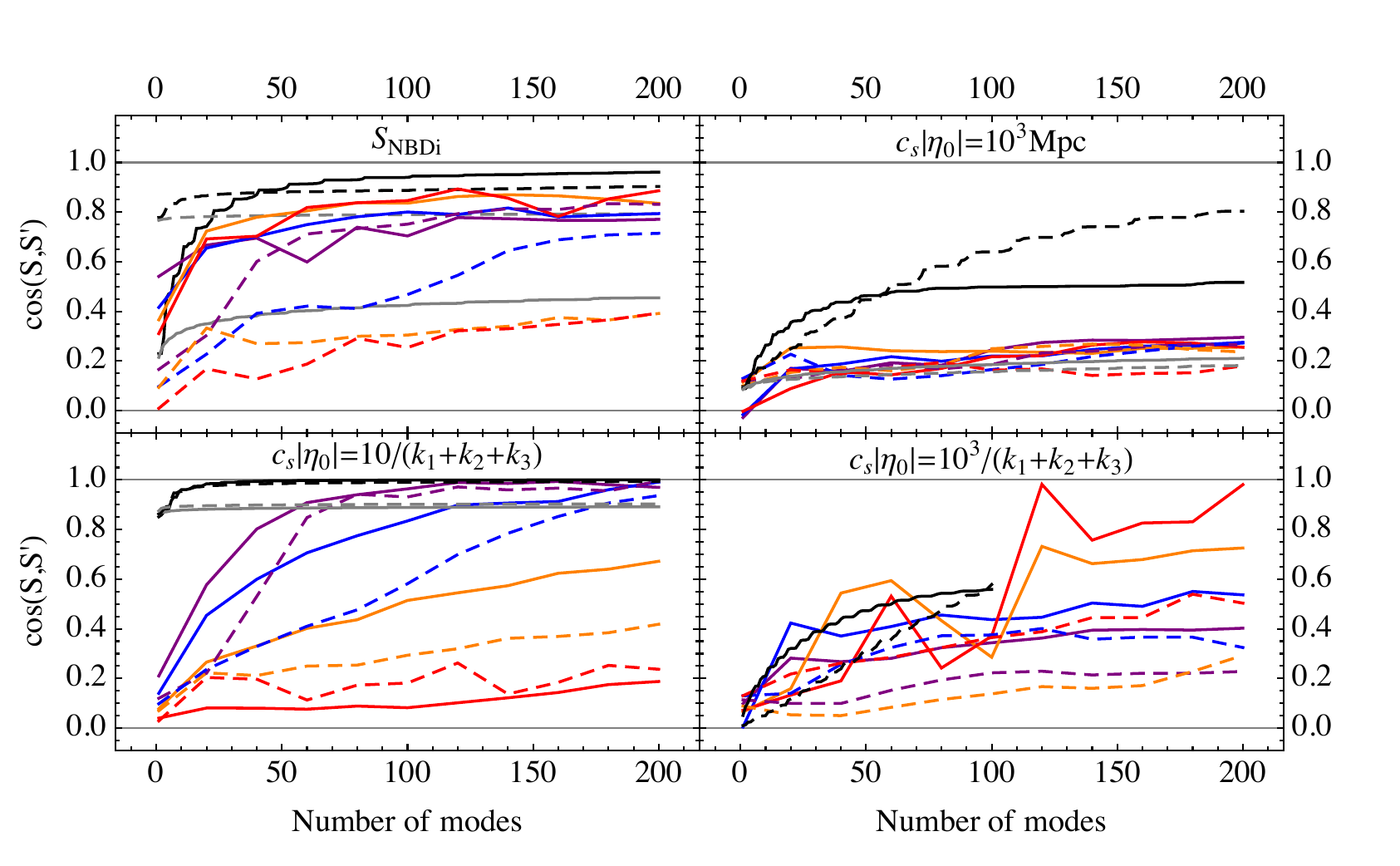}
	\caption{Comparisons of shape reconstructions for $S_{{\rm NBD1}}$ (solid) and $S_{{\rm NBD}2}$ (dashed) [upper left panel] or $S_{\dot{\zeta}^3}$ (solid) and $S_{\zeta\dot{\zeta}^2}$ (dashed) [other three panels] using the polynomial (black), Fourier (gray), and b-spline methods. 
\textit{Top row and lower left panels:} The 3D spline basis reconstructions correspond to the colored curves, where the purple, blue, orange, and red colors correspond to spline basis sets derived from using 10, 14, 22, and 40 1D spline functions in each dimension. 
\textit{Lower right panel:} The 2D spline basis reconstructions correspond to the colored curves, where the purple, blue, orange, and red colors correspond to spline basis sets derived from using 50, 100, 200, and 300 1D spline functions in each dimension.}
\label{fig:combined_cosines_figure_100}
\end{figure*}

\subsection{$S_{{\rm NBD}1}$ and $S_{{\rm NBD2}}$}

The cosines for polynomial, Fourier, and spline fits to these shapes are shown in the upper left panel of fig.\ \ref{fig:combined_cosines_figure_100}. We find that the polynomial expansion produces a better fit than the Fourier basis, and polynomial cosines typically increase slowly beyond about 100 modes for $S_{{\rm NBD}1}$ and 50 modes for $S_{{\rm NBD}2}$, indicating that lower order modes contribute most to the fits. While it is possible that increasing the number of polynomial or Fourier modes will increase the cosines further, during our analysis we found that generating large polynomial and Fourier basis sets is computationally very demanding. The separable modes are orthogonalized using a Gram-Schmidt algorithm, which is known to be numerically unstable, and the instabilities become more severe as higher order polynomials are used. On the other hand, the spline basis expansion as we have implemented it does not require orthogonalization, so in comparison a large number of modes can easily be generated and used in the separable fits. In the case of $S_{{\rm NBD}1}$ and $S_{{\rm NBD}2}$, we find that the spline expansion performs similarly well as the polynomial expansion when 200 modes are used, and better fits can be achieved by using a larger number of modes.

\subsection{$S_{\dot{\zeta}^3}$, $S_{\zeta\dot{\zeta}^2}$, and $S_{\zeta(\partial\zeta)^2}$ with $k$-independent $c_s\eta_0$}

These shapes have oscillatory features that grow with the size of the triangle (i.e. with $k_1$) and in the flattened and squeezed limits. With $c_s|\eta_0| = 10^3$ Mpc, the set of 200 polynomial modes produces a cosine of 0.52 for $S_{\dot{\zeta}^3}$ and 0.80 for $S_{\zeta\dot{\zeta}^2}$, as shown in the upper right panel of fig.\ \ref{fig:combined_cosines_figure_100}.\footnote{The shape for $S_{\zeta(\partial\zeta)^2}$ is sufficiently similar to $S_{\zeta\dot{\zeta}^2}$ that its fits are not shown separately in the figures.} While these modes produce a higher cosine than a set of 200 spline functions, better fits can be generated with a larger set of splines.

\subsection{$S_{\dot{\zeta}^3}$, $S_{\zeta\dot{\zeta}^2}$, and $S_{\zeta(\partial\zeta)^2}$ with $k$-dependent $c_s\eta_0$}

For small $\Lambda/H$, we find that the shapes are easily reconstructed with both the polynomial and spline methods using $\lesssim 200$ modes, as shown in lower left panel of fig.\ \ref{fig:combined_cosines_figure_100}. The oscillatory features are of low enough frequency that lower order polynomials are sufficient to capture most of the features of these shapes, and the spline functions also do not need to have a very fine resolution. If the initial conditions are set at $10/k_1$, rather than $10/(k_1+k_2+k_3)$, then the oscillatory features are of somewhat higher frequency; for either shape, however, the large global features still allow both the polynomial and spline bases to efficiently reconstruct these shapes. 

For cases with $k$-dependent $c_s\eta_0$ and larger values of $\Lambda/H$, the oscillations have a much higher frequency than what can be captured by polynomials, and we find that the spline reconstructions perform similarly poorly. However, the $k$-dependence of $c_s \eta_0$, whether it be $(\Lambda/H)/(k_1+k_2+k_3)$ or $(\Lambda/H)/k_1$, allows the shapes to be rewritten as functions of only two free parameters: the ratios $x \equiv k_3/k_1$ and $y \equiv k_2/k_1$. In terms of $x$ and $y$, the oscillation frequency does not increase drastically throughout the allowed parameter space, which makes it easier to generate good 2D fits. The advantage of rewriting scale-invariant shapes such as these in terms of two free parameters for computing $k$-space cosines and CMB bispectra has been discussed in earlier works \cite{Fergusson:2006pr}.

We generate the 2D fits by defining a 2D analogue of the polynomial/spline basis sets and the cosine. We find, however, that to generate the same number of polynomial modes as in the 3D case (200) requires using higher order 1D polynomials, which exacerbates the numerical issues that we encountered in the 3D case, so for the 2D fits we only use 100 polynomial modes with unit weight. We do not encounter similar numerical issues in the spline fits.

\begin{figure}[!t]
\centering
	\includegraphics[width=0.45\textwidth]{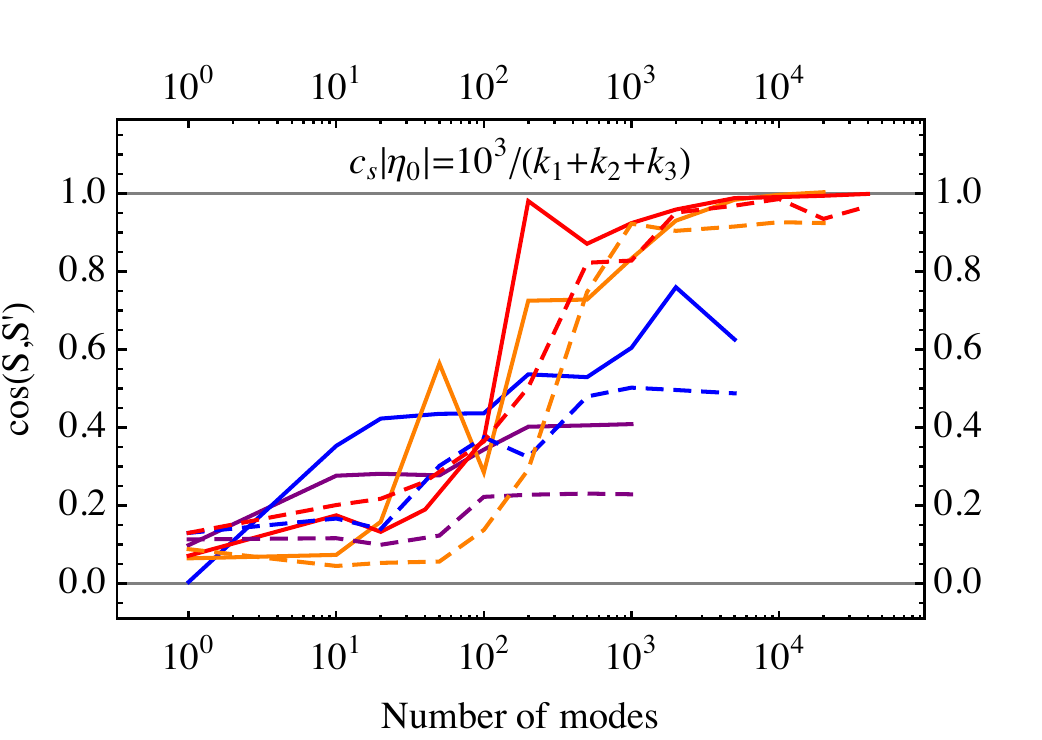}
	\caption{2D b-spline fits to $S_{\dot{\zeta}^3}$ (solid) and $S_{\zeta\dot{\zeta}^2}$ (dashed), where the purple, blue, orange, and red colors correspond to spline basis sets derived from using 50, 100, 200, and 300 1D spline functions in each dimension.}
\label{fig:extended_2d_spline_fit}
\end{figure}

We show the results for the 2D fits in the $c_s |\eta_0| = 10^3/(k_1+k_2+k_3)$ case in the lower right panel of fig.\ \ref{fig:combined_cosines_figure_100}; for $c_s |\eta_0| = 10^3/k_1$ the shapes have very similar features, with the oscillations being slightly more rapid and more difficult to represent in the latter case. 2D spline fits achieve similar cosines as 2D polynomial fits using the same number of modes ($\sim 100$), while also being able to produce higher cosines through the addition of more modes without running into numerical issues, as shown in fig.\ \ref{fig:extended_2d_spline_fit}.

\section{Discussion}
\label{sec:discussion}

If the CMB is to be used to its full potential to elucidate the details of inflation, we need to be able to both effectively characterize a wide variety of potential inflationary signatures while also ensuring that this information is accessed in a timely manner. For non-Gaussian signatures, present in the CMB bispectrum, a potential bottleneck in this process is the production of separable templates that accurately match the characteristic properties of the underlying shape. This issue is particularly acute for initial states that differ from the Bunch-Davies state, which can generically have highly oscillatory features whose characteristic scale can vary over the bispectrum configuration space.

In this work we have analyzed a variety of functions to generate templates for bispectrum shapes arising in a variety of non-Bunch-Davies scenarios. We have quantified how well different choices of separable basis functions, derived from polynomials, Fourier functions, and b-splines, can reconstruct the original non-Bunch-Davies shapes. The spline expansion method is a new alternative choice of basis, that we implement here for the first time, and can be used at a reasonable computational cost to obtain cosines that can be very close to unity. 

We find that the polynomial basis is good for describing some non-Bunch-Davies shapes with large features, and generally performs better than the Fourier basis. For the rest of the shapes we considered, assuming the low cosines will steadily increase with the addition of more modes, we find that numerical difficulties prevent us from generating enough modes to see higher cosines. For most shapes, the spline basis performs as well as the polynomial one when equal numbers of modes are chosen. The spline basis is both numerically simpler to compute, and does not require orthogonalization due to its localized nature. This allows a larger number of modes to be efficiently calculated to improve the match between the templates and the actual shapes.

The spline basis expansion method is very flexible, and there are other ways to adapt a spline basis to target specific shapes better that we have not explored here. For example, while our spline bases are derived from equally spaced knots, we note that one can create a basis using unequally-spaced knots, such that different $k$-space regions are sampled more finely than others. Non-Bunch-Davies shapes, where the features are sharply peaked and localized near flattened and squeezed configurations, can potentially be probed more efficiently by optimizing the spline basis in this way.

Finally, while our analysis takes place in primordial $k$-space, the flexibility and computational simplicity of the spline basis approach translates directly to multipole-space, where it complements existing approaches, such as the polynomial basis. Utilizing a variety of basis expansion techniques ensures that the exquisite CMB data available now, and in the future, can be used efficiently to explore the full theory space of viable inflationary scenarios.


\acknowledgments It is a pleasure to thank Donghui Jeong, Daan Meerburg, Raquel Ribeiro, and Sarah Shandera for very useful discussions. The work of J.~B. and R.~B. is supported by NASA ATP grants NNX11AI95G and NNX14AH53G, NASA ROSES grant 12-EUCLID12-0004, NSF CAREER grant 0844825 and DoE grant DE-SC0011838. J.~B. would also like to thank the Cornell Graduate School for a travel grant that facilitated the conception of this paper. R.~H. was supported in part by DOE grant DE-FG03-91-ER40682. He would also like to thank the Perimeter Institute for hospitality while this work was being completed.


\appendix
\section{Power spectrum for general initial states}
\label{app:powerspectrum}

\renewcommand{\theequation}{A\arabic{equation}}
\setcounter{equation}{0}

In this appendix we show that the result for the Green's function for general initial states in section \ref{sec:review} with a scale-dependent initial time $\eta_0(k)$ gives the correct form of the late-time power spectrum in \cite{Danielsson:2002kx,Easther:2002xe}. We start with developing a precise definition for the adiabatic vacuum at $\eta_0 \rightarrow -\infty$ or the Bunch-Davies vacuum. Let us write the field $\chi_{\vec{k}}(\eta)$ (defined via $\zeta = \frac{1}{\sqrt{2\varepsilon}} \frac{c_s}{a} \chi$) in terms of annihilation and creation operators at the time $\eta_0$, $a_{\vec{k}}(\eta_0)$ and $a_{-\vec{k}}^{\dagger}(\eta_0)$, and the mode functions $f_k^{\gtrless}(\eta)$, as
\bea
	\chi_{\vec{k}}(\eta) & = & a_{\vec{k}}(\eta_0) f_k^{>}(\eta) + a_{-\vec{k}}^{\dagger}(\eta_0) f_k^{<}(\eta) \, .
\eea
The conjugate momentum, $\pi_{\vec{k}}(\eta)$ can be obtained from the quadratic Lagrangian for $\chi_{\vec{k}}(\eta)$; at leading order this is given by
\bea
	\pi_{\vec{k}}(\eta) & = & \frac{\partial{\cal L}^{(2)}}{\partial \dot{\chi}} \ = \ \frac{\d \chi_{\vec{k}}}{\d \eta} - \frac{1}{a} \frac{\d a}{\d \eta} \chi_{\vec{k}} \, ,
\eea
We can write $\pi_{\vec{k}}(\eta)$ in terms of its corresponding mode functions $g_k^{\gtrless}(\eta)$ as
\bea
	\pi_{\vec{k}}(\eta) & = & -i \( a_{\vec{k}}(\eta_0) g_k^{>}(\eta) - a_{-\vec{k}}^{\dagger}(\eta_0) g_k^{<}(\eta) \) \, .
\eea
For Bunch-Davies modes we choose the positive frequency solution at early times, and using $a(\eta) = -1/(\eta H)$ at leading order during inflation, the mode functions are given by
\bea
	f_k^{>}(\eta) & = & \frac{1}{\sqrt{2c_s k}} e^{-i c_s k \eta} \( 1 - \frac{i}{c_s k \eta} \) \, , \\
	g_k^{>}(\eta) & = & \sqrt{\frac{c_s k}{2}} e^{-i c_s k \eta} \, .
\eea
The choice of Bunch-Davies vacuum can then be expressed as the following relationship between the field and its conjugate momentum in the infinite past,
\bea
	\pi_{\vec{k}}(\eta_0) & = & (-i c_s k) \chi_{\vec{k}}(\eta_0) \, .
\eea
Note that this does not imply that the position and momentum operators commute at all times; it is merely a statement of how they are related at $\eta_0 \rightarrow -\infty$. Equivalently, in terms of the mode function $f_k^{>}(\eta)$ we can write
\bea
	\frac{\d f_k^{>}}{\d \eta} \Big|_{\eta = \eta_0} - \frac{1}{a} \frac{\d a}{\d \eta} f_k^{>}(\eta_0) & = & (-i c_s k)  f_k^{>}(\eta_0) \, ,
\label{eq:defineadvac}
\eea
for $\eta_0 \rightarrow -\infty $. The above condition is the definition of the adiabatic vacuum in the infinite past, or equivalently the Bunch-Davies vacuum. As shown in \cite{Polarski:1995jg,Danielsson:2002kx} this choice corresponds to a minimum uncertainty state.

The prescription to choose an adiabatic vacuum at a finite initial time is to enforce the same condition in eq.\ (\ref{eq:defineadvac}) at a given $\eta_0$. This corresponds to a state which minimizes the uncertainty at $\eta = \eta_0$. Choosing the initial density matrix to be unity, but still having a non-Bunch-Davies initial state by allowing $\beta_k^{>} \ne 0$, i.e. with $f_k^{>}(\eta)$ given by eq.\ (\ref{eq:fkgtrless}) with $A_k = 0$, this leads to the following relation between the Bogoliubov coefficients,
\bea
	\frac{\beta_k^{>}}{\alpha_k^{>}} & = & \( \frac{i}{2c_s k \eta_0 + i} \) e^{-2ic_s k \eta_0} \, .
\eea
Combining this with the usual condition $\left |\alpha_k^{>} \right|^2 - \left |\beta_k^{>} \right|^2 = 1$ we find that
\bea
	\left |\alpha_k^{>} \right|^2 & = & \frac{4c_s^2 k^2 \eta_0^2 + 1}{4 c_s^2 k^2 \eta_0^2} \, .
\eea
Let us now choose $\eta_0$ to be a function of $k$ such that the physical momentum $c_s k/a(\eta_0)$ crosses some fixed high energy scale $\Lambda$ of new physics at $\eta_0$, then
\bea
	\eta_0 & = & -\frac{\Lambda}{Hc_s k} \, .
\eea
With the above equations we can write the following solution for $\alpha_k^{>}$ and $\beta_k^{>}$ (note that we are free to choose any overall phase),
\bea
	\alpha_k^{>} & = & \( \frac{2\Lambda/H - i}{2\Lambda/H} \) e^{-i\Lambda/H} \, , \\
	\beta_k^{>} & = & -\( \frac{i}{2\Lambda/H} \) e^{i\Lambda/H} \, .
\eea
For $\Lambda/H \gg 1$ we can now write the late-time power spectrum as
\bea
	{\cal P}_{\zeta}(k) & = & \frac{k^{3}}{2\pi^{2}} \frac{c_{s}^{2}}{2\varepsilon} \frac{1}{a^{2}(\eta)} f_{k}^{>}(\eta) f_{k}^{<}(\eta) \Big|_{\eta \rightarrow 0^{-}} \nonumber \\
	& = & \left. \frac{H^{2}}{8\pi^{2}\varepsilon c_{s}} \[ 1 - \frac{H}{\Lambda} \sin \( \frac{2\Lambda}{H} \) \] \right|_{c_{s}k = aH} \, , \quad \quad
\eea
which includes a scale-dependent oscillatory term. The argument of the oscillations is usually written as being proportional to $\ln (k/k_p)$, where $k_p$ is some fixed pivot scale. This can be seen by expanding $H$ around the pivot scale so that $H(k) \approx H(k_p) \[ 1 - \varepsilon (N - N_p) + \dots \] \approx H(k_p) \[ 1 - \varepsilon \ln \frac{k}{k_p} + \dots \]$, leading to logarithmic oscillations in the power spectrum.


\section{Bispectrum for general initial states}
\label{app:bispectrum}

\renewcommand{\theequation}{B\arabic{equation}}
\setcounter{equation}{0}

The functions ${\cal F}_{\dot{\zeta}^3}$, ${\cal F}_{\zeta\dot{\zeta}^2}$, and ${\cal F}_{\zeta(\partial\zeta)^2}$ in eqs.\ (\ref{eq:bispec1}) - (\ref{eq:bispec3}) are given by
\begin{widetext}
\bea
	 {\cal F}_{\dot{\zeta}^3}(p_{1},p_{2},p_{3},\eta_{0}) & = & \frac{1}{k_1 k_2 k_3} \[ -\frac{2}{K_{1}^{3}} + \frac{e^{ic_{s}K_{1}\eta_{0}}}{K_{1}} \left( \frac{2}{K_{1}^{2}} - \frac{2ic_s \eta_{0}}{K_{1}} - c_s^2 \eta_{0}^{2} \right) \] \, ,
\eea
\bea
	 & & {\cal F}_{\zeta\dot{\zeta}^2}(p_{1},p_{2},p_{3},\eta_{0}) \ = \ \frac{1}{k_1 k_2 k_3} \bigg[ \frac{- 2K_1^3 K_3^3 + K_1^2 K_2^4 + K_1 K_2^2 K_3^3}{K_1^3 K_3^6} \nonumber \\
	 & & \quad \quad \quad + \ \frac{e^{ic_{s}K_{1}\eta_{0}}}{K_1^3 K_3^6} \( 2K_1^3 K_3^3 - K_1^2 K_2^4 - K_1 K_2^2 K_3^3 + i c_s K_1^2 K_2^2 K_3^3 \eta_0 \) \bigg] \, ,
\eea
and
\bea
	 & & {\cal F}_{\zeta(\partial\zeta)^2}(p_{1},p_{2},p_{3},\eta_{0}) \ = \ \frac{1}{\( k_1 k_2 k_3 \)^3} \bigg[ \frac{K_1^6 - 3 K_1^4 K_2^2 - K_1^3 K_3^3 + 2 K_1^2 K_2^4 + 2 K_1 K_2^2 K_3^3}{K_1^3} \nonumber \\
	 & & \quad \quad \quad + \ \frac{e^{ic_{s}K_{1}\eta_{0}}}{c_s K_1^3 \eta_0} \bigg\{ i\left(K_{1}^{5} - 2K_{1}^{3}K_{2}^{2}\right) + c_{s}\left(K_{1}^{4}K_{2}^{2} + K_{1}^{3}K_{3}^{3} - 2K_{1}^{2}K_{2}^{4} - 2K_{1}K_{2}^{2}K_{3}^{3}\right)\eta_{0} \nonumber \\
	& & \quad \quad \quad \quad \quad - \ ic_{s}^{2} \left( K_{1}^{4}K_{3}^{3} - 2K_{1}^{2}K_{2}^{2}K_{3}^{3} \right)\eta_{0}^{2} \bigg\} \bigg] \, ,
\eea
\end{widetext}
where for brevity of notation we have suppressed the explicit momentum dependence of the functions $K_{1}, \ K_{2},$ and $K_{3}$,
\bea
	K_{1}(p_{1},p_{2},p_{3}) & = & p_{1} + p_{2} + p_{3} \, , \\
	K_{2}(p_{1},p_{2},p_{3}) & = & (p_{1}p_{2} + p_{2}p_{3} + p_{3}p_{1})^{1/2} \, , \\
	K_{3}(p_{1},p_{2},p_{3}) & = & (p_{1}p_{2}p_{3})^{1/3} \, .
\eea
In the next two subsections we show how one obtains the flattened and squeezed enhancements from the functions ${\cal F}$ (also see \cite{Agullo:2012cs}). We work with the simplest function ${\cal F}_{\dot{\zeta}^3}$, though the results are similar for the other two functions (or at least for their appropriate sum) as well.

\subsection{Flattened limit}

The $b_{k_1}a_{k_2}a_{k_3}$ term leads to an enhanced flattened limit ($k_1 \approx k_2 + k_3$, $k_1$ being the largest momentum mode) bispectrum. Let us first assume that $b_{k_1}$ is purely imaginary, so that the bispectrum is proportional to the imaginary part of ${\cal F}$. The corresponding ${\cal F}$ function we consider is,
\bea
	& & {\cal F}_{\dot{\zeta}^3}(-k_{1},k_{2},k_{3},\eta_{0}) \ = \ -\frac{2}{\tilde{k}_{1}^{3}k_1 k_2 k_3} \nonumber \\
	& & \quad + \ \frac{e^{ic_{s}\tilde{k}_{1}\eta_{0}}}{\tilde{k}_{1}k_1 k_2 k_3} \left( \frac{2}{\tilde{k}_{1}^{2}} - \frac{2ic_s \eta_{0}}{\tilde{k}_{1}} - c_s^2 \eta_{0}^{2} \right) \, ,
\label{eq:flatenh}
\eea
where $\tilde{k}_1 \equiv -k_1 + k_2 + k_3$. In the limit of $\tilde{k}_1 \rightarrow 0$, the exponential can be expanded as
\bea
	\lim\limits_{\tilde{k}_{1} \rightarrow 0} e^{ic_{s}\tilde{k}_{1}\eta_{0}} & = & 1 + ic_{s}\tilde{k}_{1}\eta_{0} - \frac{c_{s}^2\tilde{k}_{1}^2\eta_{0}^2}{2} - \frac{ic_{s}^3\tilde{k}_{1}^3\eta_{0}^3}{6} + \ldots \nonumber \\
\eea
Using this in eq.\ (\ref{eq:flatenh}), and noticing that any term with $\tilde{k}_1$ in the numerator goes to zero, we find that
\bea
	\lim\limits_{\tilde{k}_{1} \rightarrow 0} {\cal F}_{\dot{\zeta}^3}(-k_{1},k_{2},k_{3},\eta_{0}) & = & -\frac{i}{3(k_2 + k_3) k_2 k_3} c_s^3 \eta_0^3 \, . \nonumber \\
\eea
For fixed $\eta_0$, we can set $c_s |\eta_0| = 1/k_*$. The above limit of the three-point function is therefore enhanced (though not divergent) in the flattened limit. For $\eta_0(k)$ with large $\Lambda/H$ as well we see an enhancement in the flattened limit. If we instead assume that $b_{k_1}$ is real, then the bispectrum is still enhanced, though not in the exactly flattened limit but in a near-flattened limit.

\subsection{Squeezed limit}

Let us now look at the $a_{k_1}b_{k_2}a_{k_3}$ piece,
\bea
	& & {\cal F}_{\dot{\zeta}^3}(k_{1},-k_{2},k_{3},\eta_{0}) \ = \ -\frac{2}{\tilde{k}_{2}^{3}k_1 k_2 k_3} \nonumber \\
	& & \quad + \ \frac{e^{ic_{s}\tilde{k}_{2}\eta_{0}}}{\tilde{k}_{2}k_1 k_2 k_3} \left( \frac{2}{\tilde{k}_{2}^{2}} - \frac{2ic_s \eta_{0}}{\tilde{k}_{2}} - c_s^2 \eta_{0}^{2} \right) \, ,
\label{eq:sqenh}
\eea
where $\tilde{k}_2 \equiv k_1 - k_2 + k_3$. In the squeezed limit ($k_3 \ll k_1 \approx k_2$) we have $\tilde{k}_2 \rightarrow k_{\rm min}$, where $k_{\rm min}$ is the smallest momentum mode observable today. Using this in eq.\ (\ref{eq:sqenh}) we find that
\bea
	& & \lim\limits_{\tilde{k}_2 \rightarrow k_{\rm min}} {\cal F}_{\dot{\zeta}^3}(k_{1},-k_{1},k_{3},\eta_{0}) \ = \ -\frac{2}{k_1^2 k_{\rm min}^4} \nonumber \\
	& & \quad \quad + \ \frac{e^{ic_{s}k_{\rm min}\eta_{0}}}{k_1^2 k_{\rm min}^2} \left( \frac{2}{k_{\rm min}^2} - \frac{2ic_s \eta_{0}}{k_{\rm min}} - c_s^2 \eta_{0}^{2} \right) \, . \quad \quad
\eea
For fixed $\eta_0$ and in the limit of $k_{\rm min} \gg k_*$, the $c_s^2 \eta_0^2$ term gives the largest contribution. This term is multiplied with a highly oscillatory function though, and averaging over the large argument of the cosine (real part of the exponential) we expect its contribution to vanish. The leading order contribution is then proportional to $1/(k_1^2 k_{\rm min}^4)$, which shows a strong squeezed limit enhancement. In the limit of $k_{\rm min} \gtrsim k_*$ or for $\eta_0(k)$ this argument no longer holds and we may or may not see enhancements.\footnote{In our fits with fixed $\eta_0$ in section \ref{sec:splinesNBD} we took $k_*$ to be similar to $k_{\rm min}$ ($\sim 10^{-3}$ Mpc$^{-1}$) which is what the Planck team had used in \cite{Ade:2013ydc}.}


\section{B-splines fitting algorithm}
\label{app:splinefit}

\renewcommand{\theequation}{C\arabic{equation}}
\setcounter{equation}{0}

In this appendix, we build on the discussion of section \ref{sec:splines} to describe in more detail the b-spline fitting algorithms we have used, and illustrate b-spline fitting examples for a simple 1D function and a 2D representation of the scale-invariant enfolded template. Snippets of Mathematica 10 codes we have implemented are shown here, and the same algorithms were first constructed and explicitly shown as Matlab code in \cite{Eilers:2006}.

Our first example considers a spline fit to a simple 1D function $f(x)$, with $0 \leq x \leq 1$. To generate a fit, we first make two choices: a choice of basis and a choice of data points to fit. The spline basis is determined by a choice of $k+1$ equidistant knots, $\{ x_0, x_1, \dots, x_k \}$, at which each basis function's degree $q$ polynomial pieces will be joined. The b-splines in general do not have to be generated using constant knot intervals, but for simplicity we always start with equidistant knots, and additionally include $q$ extra knots at each of $x=0$ and $x=1$ to produce a ``clamped'' basis. Without these extra knots, the generated basis sets do not have splines with non-zero amplitudes at $x=0$ and $x=1$, and it will be difficult to fit functions that are non-zero at the endpoints.

Given these inputs, existing codes, such as Mathematica's \texttt{BSplineBasis} function, can recursively generate a basis of $k+q$ b-splines such that each one is spanned by $q+2$ knots, made up of $q+1$ polynomial pieces, with derivatives continuous up to order $q-1$. In addition, the sum of all b-spline amplitudes at any $x$ is unity. For example, the b-splines in fig.\ \ref{fig:sample_1d_bsplines} are easily generated by choosing $q=3$ and $k=3$ such that the knots vector is \texttt{knots = \{0,0,0,0,1/3,2/3,1,1,1,1\}}, and executing \texttt{BSplineBasis[\{q,knots\},i,x]}, where $i$ is an integer $0 \leq i \leq k+q-1$ identifying each particular b-spline. In this work, we vary the number and widths of the b-splines by varying $k$, but always keep the degree fixed to $q=3$.

Next, the choice of data points depends on how finely we wish to sample $f(x)$. We would like to choose a data set consisting of $M$ data pairs, $(x_i, f_i)$, where $f_i \equiv f(x_i)$, such that our data resolves any potentially fine features in the function we would like to fit, without including so many extra data points that our numerical calculation becomes intractable. The final fit should ultimately be insensitive to the sampling we have chosen. Again, for simplicity, we always use equidistant sampling points $x_i$ in our analysis, but vary the density of sampling points by changing $M$.

The spline fit then approximates the original function as
\bea
	f'(x) & = & \sum_{n=0}^{N-1} \alpha_n B_n(x) \, ,
\eea
where the expansion coefficients $\{\alpha_n\}$ are solved for by minimizing the least-squares function,
\bea
	{\rm LS} & = & \sum_{i=1}^M \left( f_i - \sum_{n=0}^{N-1} \alpha_n B_n(x_i) \right)^2 \, .
\label{1DLS}
\eea
This requires solving the linear system of equations given by $B^T B \vec{\alpha} = B^T \vec{f}$, where $B$ is an $M \times N$ matrix containing the values of $B_n(x_i)$, and is easily performed with algorithms such as Mathematica's \texttt{LinearSolve}.

We note one extension of b-splines, called p-splines, that aims to avoid overfitting a set of input data by imposing smoothness on the resulting fit. Short for ``penalty b-splines'', in the p-spline method, the fit's expansion coefficients are determined by both the choice of basis and the choice of input data, plus a choice of penalty function that generally disfavors fits with large differences between coefficients of neighboring b-splines \cite{Eilers:1996}. In this context one would instead minimize
\bea
	{\rm LS} & = & \sum_{i=1}^M \left( f_i - \sum_{n=0}^{N-1} \alpha_n B_n(x_i) \right)^2 + \lambda \sum_{j=k}^{N-1} \left( \Delta^k \alpha_j \right)^2 \, , \nonumber \\
\eea
where $\lambda$ is a constant that controls the smoothness of the fit and $k$ is the order of the penalty, a typical choice being $k=2$, such that $\Delta^2 \alpha_j = \alpha_j -2 \alpha_{j-1} + \alpha_{j-2}$. The use of a penalty is optional, and its main purpose in the context of data fitting is to avoid fitting any noisy features in the data. Further, if there are not sufficiently many data points sampling $f(x)$, with many more splines than data points, then without a penalty the fits may display spurious features, as we will deliberately try to show in the 1D example that follows.

\begin{figure}[!t]
\centering		
	\includegraphics[width=0.45\textwidth]{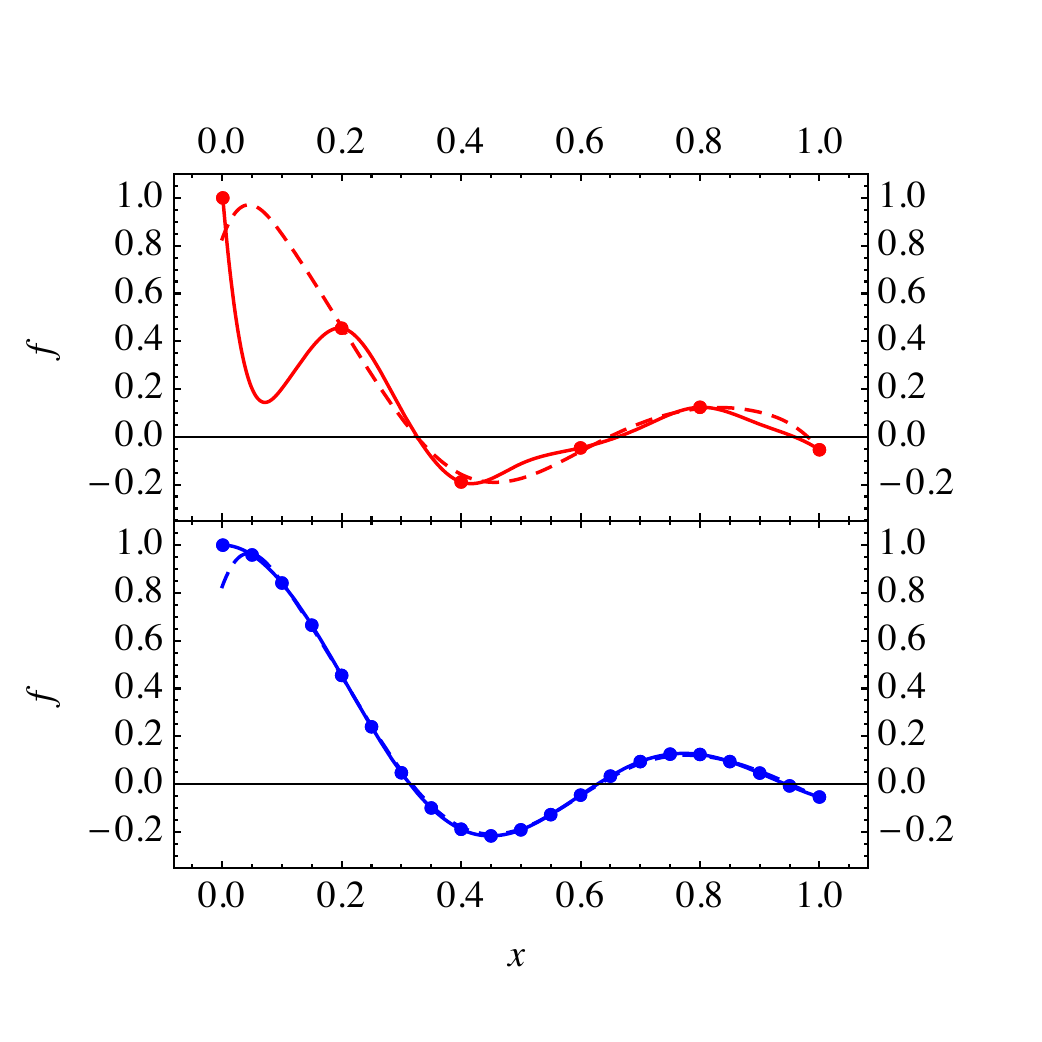}
	\caption{Example 1D b-spline fits to $f(x) = \sin(10x)/(10x)$. All fits shown have used a basis of 13 splines. In the upper panel, the fits have been computed using 6 data points, sampled at $x_i = \{0,0.2,0.4,0.6,0.8,1\}$. Attempting to construct such a fit without smoothing [solid red] causes the splines to produce spurious features, especially at smaller values of $x$, while introducing a penalty and a small amount of smoothing [dashed red], $\lambda = 0.1$, restores the fit to a reasonable representation of the true function. The lower panel is the same fit, except with 21 data points, sampled at $x_i = \{0,0.05,0.1,\dots,0.9,0.95,1\}$. In this case, the fits with [dashed blue] and without [solid blue] smoothing are very similar.}
\label{fig:spline_1D_example}
\end{figure}

In fig.\ \ref{fig:spline_1D_example} we show fits to $f(x) = \sin(10x)/(10x)$ with different choices of data points and smoothing parameter $\lambda$. We have fixed the knots at $\{ 0,0,0,0,0.1,0.2, \dots, 0.8,0.9,1,1,1,1\}$, yielding a basis of 13 b-splines. We find, as illustrated in the figure, that we can achieve good fits without introducing extra smoothing through a non-zero value of $\lambda$, as long as we fit to enough data points. So we now continue to an example of fitting a primordial shape in two dimensions, without a penalty.

To illustrate the b-spline fitting algorithm in two dimensions, we construct a basis of 2D b-splines and use it to fit the scale-invariant enfolded template,
\bea
	S_{\rm enf}(x,y) & = & \frac{1}{xy} \, (1 - x - y - x^2 - y^2 + x^3 + y^3 \nonumber \\
	& & \quad \quad \ \ - \ x^2 y - x y^2 + 3 x y) \, ,
\eea
where $x \equiv k_3/k_1$ and $y \equiv k_2/k_1$. As in the 1D case, the inputs to the fitting algorithms are made up of a choice of basis and a set of data points.  The basis is specified by a choice of polynomial degree and a sequence of knots in each of the two dimensions, $x$ and $y$. In our particular application, since we are aiming to fit shape functions that are symmetric in their wavenumber arguments, we only specify the knots and degree in one dimension, and use the same b-spline basis for the additional second dimension. The 2D b-splines are then made of products of any two 1D splines, for example, $B_n(x)B_m(y)$. The data are given by $(x_i,y_j,S(x_i,y_j))$ and stored in $Y_{ij} = S(x_i,y_j)$, where again some care must be taken in the choice of sampling, which must be dense enough to capture any small features such as oscillations that we would like to capture in the resulting fit.

The 2D analogue of the least-squares function in eq.\ \eqref{1DLS} is 
\bea
	{\rm LS} & = & \sum_{i=1}^M \sum_{j=1}^M \left( Y_{ij} - \sum_{m=0}^{N-1}\sum_{n=0}^{N-1} \alpha_{mn} B_m(x_i) B_n(y_j) \right)^2 \, . \nonumber \\
\eea
To turn the problem of solving for the expansion coefficients $\alpha_{mn}$ into a linear system, we create a regression basis $C$ from the $M \times N$ b-spline basis matrices in each dimension, $B_1$ and $B_2$, which in our case are equal. We define
\bea
	C & = & (B_2 \otimes e_L^T) \odot (e_K^T \otimes B_1) \equiv B_1 \square B_2 \, ,
\eea
where $\otimes$ is the Kronecker product, $\odot$ is an element-by-element multiplication, the second equality defines the $\square$ operation, and $e_L$ and $e_K$ are vectors of 1's with length $L$ each. In Mathematica, we define the $\square$ operation as \texttt{box}:
\begin{lstlisting}
box[B1_,B2_]:=Module[{K,L,eK,eL},
K=Length[B1[[1,All]]];
L=Length[B2[[1,All]]];
eK=ConstantArray[1,{K}];
eL=ConstantArray[1,{L}];
KroneckerProduct[B2,{eL}]
  *KroneckerProduct[{eK},B1]
]
\end{lstlisting}
Then by stacking the columns of the coefficients array $\alpha_{mn}$ and the data array $Y_{ij}$ to get vectors $\vec{\beta}$ and $\vec{y}$ respectively, the task of finding a solution for the coefficients is once again reduced to solving a linear system of equations given by $C^T W C \vec{\beta} = C^T W\vec{y}$. Here $W$ is a matrix containing weights, which may be different for each data point, but for simplicity we restrict ourselves to using a weight of unity for all of our data.

For modest amounts of data and numbers of b-splines, one can quickly solve for the coefficients in this straightforward way. For large data sets and numbers of b-splines, however, this approach becomes computationally cumbersome due to the large size of $C$. While it is still possible to numerically solve for the coefficients $\alpha_{mn}$ using a low-level language like C(++) or Fortran, we have instead used algorithms developed for higher level languages, such as Matlab in \cite{Eilers:2006}, using only vector and matrix operations. This has the benefit of being easier to implement, while still being able to sidestep much of the memory storage and speed issues typical of a more brute-force approach in a high-level language. Instead of starting with a calculation of $C$ in the brute-force approach, the algorithm from \cite{Eilers:2006} that we have adopted computes $C^T W C$ and $C^T W \vec{y}$ using only $B$. We refer the reader to \cite{Eilers:2006} for a detailed discussion of how the method itself is devised and constructed, or to see the equivalent Matlab code, and present here an implementation of the b-spline fitting algorithms in Mathematica.

The normal equations can be efficiently constructed and solved, given an input of data in $Y$ and information about the data sampling and b-spline basis in $B$:
\begin{lstlisting}
get2dfit[Y_,B_]:=Module[{m,n,W,R,r,
   F,a,A},
m=Length[Y[[1,All]]];
n=Length[B[[1,All]]];
W=ConstantArray[1,{m,m}]; 
R=Transpose[B].(W*Y).B;
r=ArrayReshape[R,{n*n,1}]; 
F=Transpose[box[B,B]].W.box[B.B];
F=ArrayReshape[F,{n,n,n,n}]; 
F=TensorTranspose[F,Cycles{{3,2}}]]; 
F=ArrayReshape[F,{n*n,n*n}]; 
a=LinearSolve[F,r];
A=ArrayReshape[a,{n,n}]
]
\end{lstlisting}

After the matrix of coefficients is solved for, we construct the final fit through two steps. First, we map the coefficients output as $A$ from \texttt{get2dfit} to a new set of coefficients that corresponds to a 2D basis of splines which is symmetric in its two arguments, so that each 2D basis mode is a sum of up to two terms: $B_i(x)B_j(y) + B_j(x)B_i(y)$. Second, in building up the fit, mode by mode, we start with the modes that contribute most to the fit. Since the b-splines in a choice of basis have similar shapes and amplitudes, we use the magnitude of the $A_{ij}$ coefficient as a proxy for gauging how much any particular b-spline contributes to a fit's overall cosine with the original shape. This motivates building up a fit by adding in modes, starting with those that have the largest $|A_{ij}|$. The cosine is then computed in the usual way, through an inner product over $(x,y)$-space between the original shape $S_{\rm enf}$ and the fit.

\begin{figure}[!t]
	\centering		
	\includegraphics[width=0.45\textwidth]{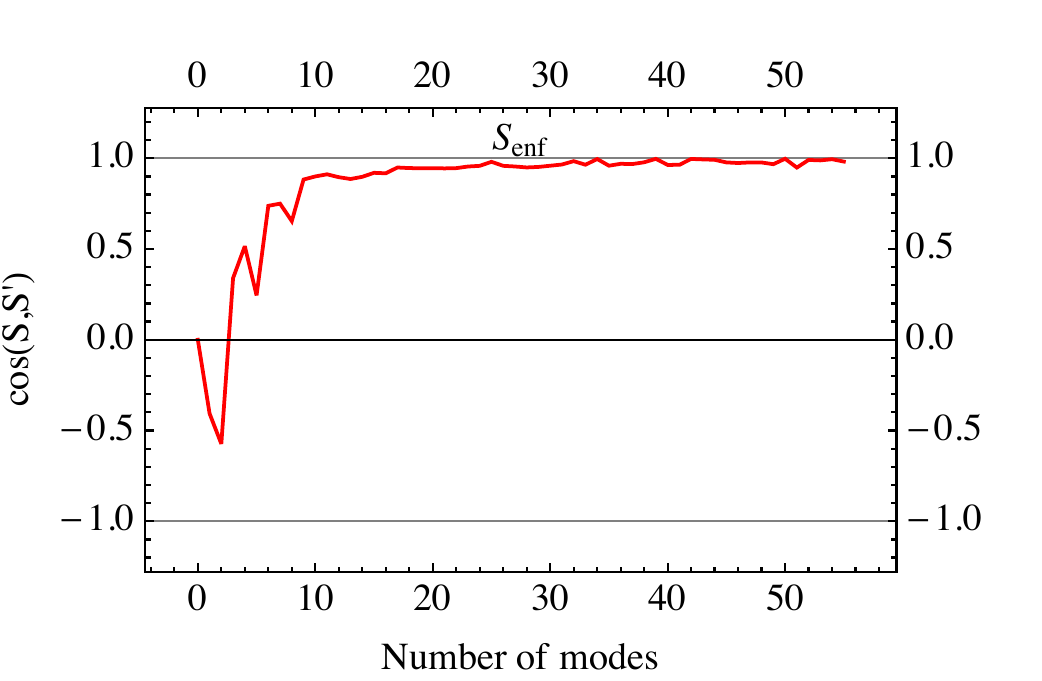}
	\caption{Cosines between the enfolded shape and the 2D b-spline fits generated by a basis with 10 splines per dimension.}
	\label{fig:spline_2D_example}
\end{figure}

\begin{figure}[!h]
	\centering		
	\includegraphics[width=0.45\textwidth]{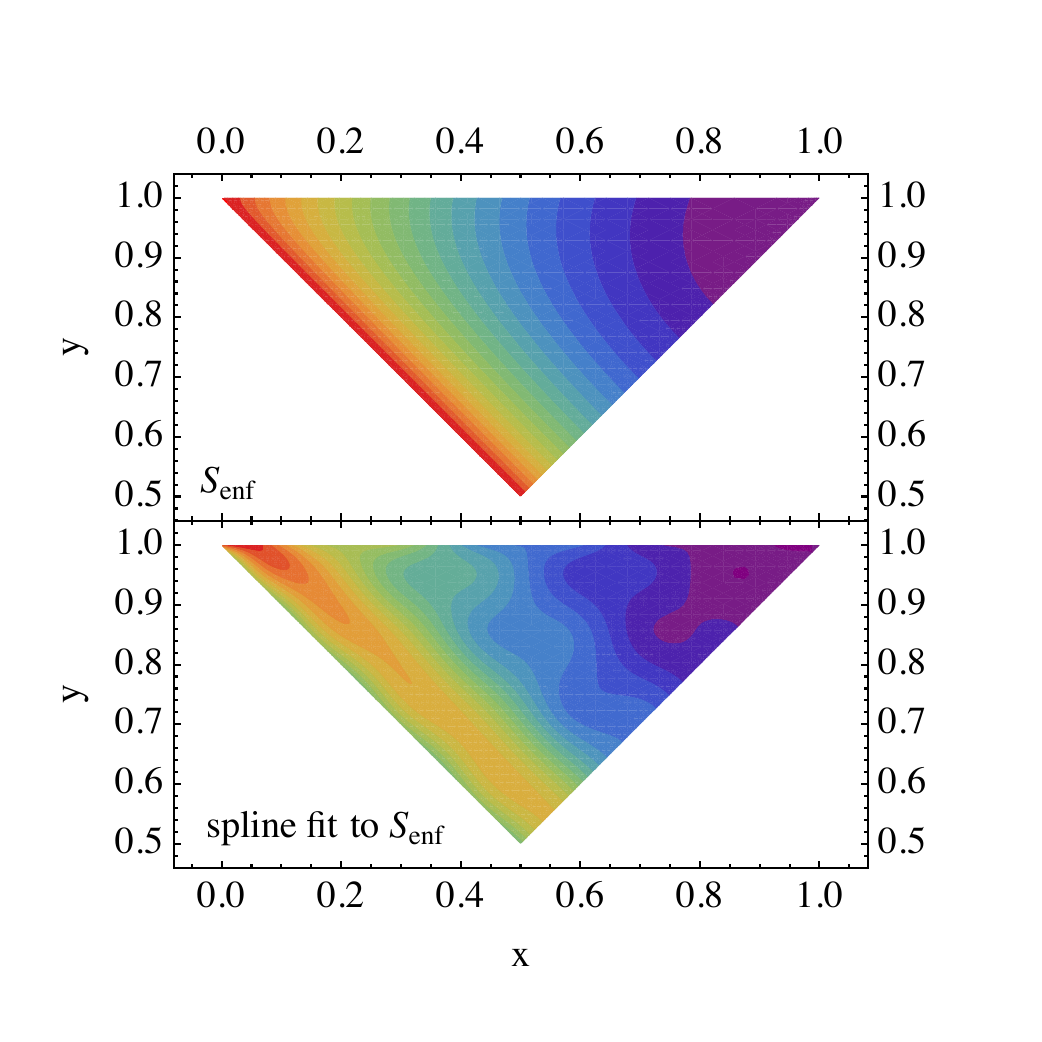}
	\caption{The original template $S_{\rm enf}(x,y)$ [upper panel] and its b-spline fit using a set of 55 modes [lower panel]. The amplitudes range from 0 (violet) to 1 (red) in both panels.}
	\label{fig:spline_2D_visual}
\end{figure}

As an example, we use a basis of 10 splines in each dimension constructed by choosing $k=7$ and $q=3$, and use as our data set a grid of uniformly spaced $(x,y)$ values from taking 50 samples in each dimension, to compute the matrix of coefficients $A_{ij}$, using the algorithms \texttt{box} and \texttt{get2dfit} above. The total number of symmetric modes is then 55, and the modes are ordered by their corresponding values of largest to smallest $|A_{ij}|$ to produce the cosines in fig.\ \ref{fig:spline_2D_example}. A visual comparison of the full fit using 55 modes and the original enfolded template is given in fig.\ \ref{fig:spline_2D_visual}.

For our 3D fits, the fitting method is the same: a choice of b-spline basis and data set make up the inputs to the algorithm, which returns an array $A_{mnp}$ containing the expansion coefficients that approximate the input shape as $S'(k_1,k_2,k_3) = \sum_{mnp} A_{mnp} \, B_m(k_1) B_n(k_2) B_p(k_3)$.  However, due to the higher dimensionality of the problem, we must introduce a new function, \texttt{rho}, to generalize the matrix product to the product of a matrix and a 3D array. Below, we show Mathematica code for \texttt{rho[A,B,p]}, which computes the normal matrix product between rows of $A$ and the $p^{\rm th}$ column of $B$, resulting in a product, $C$, which has the same dimensions as $B$:
\begin{lstlisting}
rho[A_,B_,p_]:=Module[{sa,sb,n,ip,
  cycles,sbip,prodsbip,tempB,C},
sa=Dimensions[A];
sb=Dimensions[B];
n=Length[sb];
ip=Join[Range[p+1,n],Range[1,p-1]];
Which[
  p==1,cycles=Cycles[{}],
  p==2,cycles=Cycles[{{1,3,2}}],
  p==3,cycles=Cycles[{{1,2,3}}]];
tempB=TensorTranspose[B,cycles];
sbip=sb[[ip]];
prodsbip=Product[sbip[[i]],
  {i,1,Length[sbip]}];
tempB=ArrayReshape[tempB,{sb[[p]],
  prodsbip}];
C=Transpose[A].tempB;
C=ArrayReshape[C,Join[{sa[[2]]},
  sb[[ip]]]];
C=TensorTranspose[C,
  InversePermutation[cycles]]
]
\end{lstlisting}
Given this definition of \texttt{rho}, the 3D b-spline fit coefficients are calculated using \texttt{get3dfit}:
\begin{lstlisting}
get3dfit[Y_,B_]:=Module[{m,n,W,F,R,A},
m=Length[Y[[1,1,All]]];
n=Length[B[[1,All]]];
W=ConstantArray[1,{m,m,m}];
F=rho[box[B,B],W,1];
F=rho[box[B,B],F,2];
F=rho[box[B,B],F,3];
R=rho[B,Y*W,1];
R=rho[B,R,2];
R=rho[B,R,3];
F=ArrayReshape[F,{n,n,n,n,n,n}];
F=TensorTranspose[F,
  Cycles[{{3,2,4,5}}]];
F=ArrayReshape[F,{n^3,n^3}];
A=LinearSolve[F,
  ArrayReshape[R,{n^3,1}]];
A=ArrayReshape[A,{n,n,n}]
]
\end{lstlisting}


\bibliographystyle{apsrev}
\bibliography{references}

\end{document}